\documentclass[12pt]{amsart}
\usepackage{amsmath}
\usepackage{amsthm}
\usepackage{amsfonts}
\usepackage{graphicx}
\usepackage{amssymb} %for \geqslant
\usepackage{bbold}
\newenvironment{nouppercase}{
  
  \renewcommand{\uppercasenonmath}[1]{}}{}

\begin{document}

\title[The fast non-commutative sharp drop]
{The fast non-commutative sharp drop}
\author[Jens Hoppe]{Jens Hoppe}
%\date{18 Feb 2023}
\address{Braunschweig University, Germany \& IHES, France}
\email{jens.r.hoppe@gmail.com}

\begin{abstract}
An exact GH \cite{1} membrane matrix model solution is given that corresponds to the world volume swept out by a fast moving axially symmetric drop with a sharp tip.
\end{abstract}

\begin{nouppercase}
\maketitle
\end{nouppercase}
\thispagestyle{empty}
%%%%%%%%%%%%%%%%%%%%%%%%%%%%%%%%%%%%%%%%%%%%%%%%%%%%%%%%%%%%%%%%%%%%%%%%%%%%%
%PAGE 1 PAGE 1 PAGE 1 PAGE 1 PAGE 1 PAGE 1 PAGE 1 PAGE 1 PAGE 1 PAGE 1 PAGE 1 
%%%%%%%%%%%%%%%%%%%%%%%%%%%%%%%%%%%%%%%%%%%%%%%%%%%%%%%%%%%%%%%%%%%%%%%%%%%%%
\noindent
Exact solutions of non-linear differential equations, as useful as they can be, are generally difficult to get, and for the case of the membrane matrix model,
\begin{equation}\label{eq1} 
\begin{array}{l}
\ddot{X}_i = -\sum_{j=1}^d\big[ [X_i, X_j], X_j \big],\\[0.25cm]
\sum_i [X_i, \dot{X}_i] = 0, \quad X_i^\dagger = X_i
\end{array}
\end{equation}
(the $X_i \, (i = 1, \ldots d)$ being time-dependent hermitean matrices) not many are known. In \cite{2} it was shown that (time-independent) representations of $so(1, d=2)$ can be used to give non-commutative analogues of the `minimal' (zero-mean curvature) time-like infinitely extended 3-manifolds defined by \cite{3} 
%%%%%%%%%%%%%%%%%%%%%%%%%%%%%%%%%%%%%%%%%%%%%%%%%%%%%%%%%%%%%%%%%%%%%%%%%%%%%
%PAGE 2 PAGE 2 PAGE 2 PAGE 2 PAGE 2 PAGE 2 PAGE 2 PAGE 2 PAGE 2 PAGE 2 PAGE 2 
%%%%%%%%%%%%%%%%%%%%%%%%%%%%%%%%%%%%%%%%%%%%%%%%%%%%%%%%%%%%%%%%%%%%%%%%%%%%%
\begin{equation}\label{eq2} 
(t^2+x^2+y^2-z^2)(t+z)^2 = C < 0.
\end{equation}
A non-commutative analogue of the (for fixed $t$) compact variant
\begin{equation}\label{eq3} 
(t^2-x^2-y^2-z^2) = C(t+z)^6 > 0
\end{equation}
was described in \cite{4}.\\
In this note I would like to give exact solutions to \eqref{eq1} that correspond to the one-parameter family of zero-mean-curvature $\mathfrak{M}_3 \subset \mathbb{R}^{1,3}$ defined by 
\begin{equation}\label{eq4} 
(t^2+x^2+y^2-z^2) - 6C\sqrt{x^2+y^2}(t+z)^2 + 3C^2(t+z)^4 = 0
\end{equation}
which for $C<0$ describes a fast moving ($ z = -t,\, r=0$ with velocity $-1$), for each fixed (say, positive) $t > \frac{-1}{9C} > 0$ `compact' axially symmetric drop with a sharp tip at $z = -t-\delta t < -t$,
\begin{equation}\label{eq5} 
3C^2(\delta t)^3 - (\delta t) - 2t = 0;
\end{equation}
note that $\mathfrak{M}_3^{C<0}$ may also be written as the set of points where
\begin{equation}\label{eq6} 
v(x) := \frac{r}{(t+z)^2}\big( 1- \sqrt{\frac{2}{3} + \frac{z^2-t^2}{3r^2}} \big) = C < 0,
\end{equation}
%%%%%%%%%%%%%%%%%%%%%%%%%%%%%%%%%%%%%%%%%%%%%%%%%%%%%%%%%%%%%%%%%%%%%%%%%%%%%
%PAGE 3 PAGE 3 PAGE 3 PAGE 3 PAGE 3 PAGE 3 PAGE 3 PAGE 3 PAGE 3 PAGE 3 PAGE 3 
%%%%%%%%%%%%%%%%%%%%%%%%%%%%%%%%%%%%%%%%%%%%%%%%%%%%%%%%%%%%%%%%%%%%%%%%%%%%%  
resp.
\begin{equation}\label{eq7} 
\begin{split}
r(t,z) & = (t+z)^2(3C + \sqrt{6C^2 + f(z;t)})\\[0.15cm]
f(z;t) & = \frac{z-t}{(z+t)^3}, \qquad f'(z) = \frac{4t-2z}{(t+z)^4}
\end{split}
\end{equation}
resp. (letting $r= \frac{\tilde{r}}{-C}$, $t= \frac{\tilde{t}}{-C}$, $z = \frac{\tilde{z}}{-C}$; this scale-`symmetry' could of course be used already above to effectively put $C = -1$; in particular: $v(x)$ should be a solution of the level-set minimality condition without restricting to $\mathfrak{M}_3$)
\begin{equation}\label{eq8} 
\begin{array}{l}
\tilde{r}(\tilde{t}, \kappa := \frac{t+z}{-C}) = 3\kappa^2(\sqrt{\frac{2}{3} + g(\kappa,\tilde{t})}-1) \\[0.25cm]
g(\kappa) = \frac{1}{9}(\frac{1}{\kappa^2}-\frac{2\tilde{t}}{\kappa^3}), \quad g'(\kappa) = \frac{1}{9}(\frac{6\tilde{t}-2\kappa}{\kappa^4})
\end{array}
\end{equation}
(for positive $\kappa$, $g$ has a maximum, for $\kappa = 3\tilde{t}$, i.e. $z = 2t$, with value $\frac{1}{3}\cdot \frac{1}{81\tilde{t}^2}$; which is the reason that for $\tilde{t} > \frac{1}{9}$ only negative $\kappa$ can make \eqref{eq8} positive; $g'$ being positive there restricts $\kappa$ to the interval $[-\kappa_M,0)$, with $3\kappa^3_M - \kappa_M -2\tilde{t} = 0$, $g(-\kappa_M) = \frac{1}{3}$, which has a unique positive solution $\kappa_M > 0$ for any given $\tilde{t}>0$).
%%%%%%%%%%%%%%%%%%%%%%%%%%%%%%%%%%%%%%%%%%%%%%%%%%%%%%%%%%%%%%%%%%%%%%%%%%%%%
%PAGE 4 PAGE 4 PAGE 4 PAGE 4 PAGE 4 PAGE 4 PAGE 4 PAGE 4 PAGE 4 PAGE 4 PAGE 4 
%%%%%%%%%%%%%%%%%%%%%%%%%%%%%%%%%%%%%%%%%%%%%%%%%%%%%%%%%%%%%%%%%%%%%%%%%%%%% 
For $z\nearrow -t$
\begin{equation}\label{eq9} 
\begin{split}
\frac{\partial r}{\partial z} & = 2(t+z)(3c + \sqrt{\hphantom{xx}}) + \frac{2t-z}{(t+z)^2\sqrt{\hphantom{xx}}}\\
& \approx -2 \sqrt{\frac{z-t}{z+t}} + \frac{2t-z}{\sqrt{z^2 - t^2}}\\
& = \frac{z}{\sqrt{z^2-t^2}} \searrow -\infty
\end{split}
\end{equation}
(as it should, for a closed surface to be round, at the top) while for $z \searrow -t -\delta t$
\begin{equation}\label{eq10} 
\begin{split}
\frac{\partial r}{\partial z} & \approx -2\delta t(3c + \sqrt{\hphantom{xx}}) + \frac{t + \delta t}{(\delta t)^2\sqrt{\hphantom{xx}}}\\
& = \frac{1}{(\delta t)^2\sqrt{\hphantom{x}}}\big( t + \delta t - 2(\delta t)^3(\underbrace{3C\sqrt{\hphantom{xx}} + \sqrt{\hphantom{xx}}^2)}_{\rightarrow 0} \big)\\
& \rightarrow \frac{t+\delta t}{(\delta t)^2 (-3C)} = \text{finite} > 0,
\end{split}
\end{equation}
meaning that at the bottom the drop is sharp, rather than round (though getting less and less sharp as $t \rightarrow \infty$); to show that $\frac{\partial r}{\partial z}$ is a monotonically decreasing function in the interval $[-t-\delta t, -t)$ (so, in particular, will have only one zero, at height $\hat{z}$ - where the radius $r$ becomes maximal) is cumbersome (though, at least for large enough $t$, presumably true).\\
%%%%%%%%%%%%%%%%%%%%%%%%%%%%%%%%%%%%%%%%%%%%%%%%%%%%%%%%%%%%%%%%%%%%%%%%%%%%%
%PAGE 5 PAGE 5 PAGE 5 PAGE 5 PAGE 5 PAGE 5 PAGE 5 PAGE 5 PAGE 5 PAGE 5 PAGE 5 
%%%%%%%%%%%%%%%%%%%%%%%%%%%%%%%%%%%%%%%%%%%%%%%%%%%%%%%%%%%%%%%%%%%%%%%%%%%%% 
As realized already more than 40 years ago \cite{1}, axially symmetric time-like minimal 3 manifolds can be described by solutions $R(\tau,\mu)$ of the simple-looking PDE
\begin{equation}\label{eq11} 
\ddot{R} = R(RR')';
\end{equation}
$x_1 = R\cos \psi$ and  $x_2 = R\sin \psi$ will then satisfy
\begin{equation}\label{eq12} 
\begin{split}
\ddot{x}_i & = \sum_{j=1}^2 \left\lbrace \{x_i, x_j \}, x_j \right\rbrace = \Delta x_i \\
\sum_j \{ x_j, \dot{x}_j \} & := \sum_j \big( \frac{\partial x_j}{\partial \mu}\frac{\partial \dot{x}_j}{\partial \psi} - \frac{\partial x_j}{\partial \psi}  \frac{\partial \dot{x}_j}{\partial \mu}\big)  = 0
\end{split}
\end{equation}
with respect to a $(\mu, \psi)$ Poisson bracket, and $\zeta = t-z$, (re)constructed via 
\begin{equation}\label{eq13} 
\begin{split}
\zeta' & = \dot{R}R' \\
\dot{\zeta} & = \frac{1}{2}(\dot{R}^2 + R^2R'^2)
\end{split}
\end{equation}
will also be in the kernel of the non-linear wave-operator $\square = \partial^2_t - \Delta$,
\begin{equation}\label{eq14} 
\ddot{\zeta} = \sum_j \left\lbrace \{ \zeta, x_i \}, x_i \right\rbrace \; (= (R^2 \zeta')'). 
\end{equation}
The solution of \eqref{eq13} corresponding to the `fast moving sharp drop' \eqref{eq4} is (with $\varepsilon = 4C \leq 0$, cp.\cite{5})
\begin{equation}\label{eq15} 
\begin{split}
R & = \sqrt{2} \frac{\mu}{\tau} + \varepsilon \tau^2 \\
\zeta & = \varepsilon^2 \tau^3 - \frac{\mu^2}{\tau^3} + 2\sqrt{2} \varepsilon \mu,
\end{split}
\end{equation}
%%%%%%%%%%%%%%%%%%%%%%%%%%%%%%%%%%%%%%%%%%%%%%%%%%%%%%%%%%%%%%%%%%%%%%%%%%%%%
%PAGE 6 PAGE 6 PAGE 6 PAGE 6 PAGE 6 PAGE 6 PAGE 6 PAGE 6 PAGE 6 PAGE 6 PAGE 6 
%%%%%%%%%%%%%%%%%%%%%%%%%%%%%%%%%%%%%%%%%%%%%%%%%%%%%%%%%%%%%%%%%%%%%%%%%%%%% 
i.e
\begin{equation}\label{eq16} 
\begin{split}
x_1 & = (\sqrt{2}\frac{\mu}{\tau} + \varepsilon \tau^2)\cos \psi = \frac{\sqrt{2}}{\tau} \big(\bar{x}_1 := (\mu + \frac{\varepsilon \tau^3}{\sqrt{2}}) \cos \psi \big) \\
x_2 & = (\sqrt{2}\frac{\mu}{\tau} + \varepsilon \tau^2)\sin \psi = \frac{\sqrt{2}}{\tau} \big(\bar{x}_2 := (\mu + \frac{\varepsilon \tau^3}{\sqrt{2}}) \sin \psi \big)
\end{split}
\end{equation}
will solve \eqref{eq12}. Defining
\begin{equation}\label{eq17} 
\bar{x}_0 := \{ \bar{x}_1, \bar{x}_2 \} = \mu + \frac{\varepsilon \tau^3}{\sqrt{2}} =: \mu + e
\end{equation}
one has (with $\varepsilon_{012} = +1$, and raising indices with respect to $\eta^{\alpha \beta} = $ diag$(1,-1,-1)$ metric) 
\begin{equation}\label{eq18} 
\{ \bar{x}_{\alpha}, \bar{x}_{\beta}\} = \varepsilon_{\alpha \beta} \,^\gamma \bar{x}_{\gamma}, \quad -q := \bar{x}_0^2 - \bar{x}_1^2 - \bar{x}_2^2 = 0,
\end{equation}
which together with the equally crucial property $\ddot{x}_i = \frac{2}{\tau^2}x_i$ almost trivially implies \eqref{eq12}. Compared to previously discussed solutions (e.g. \eqref{eq2} in \cite{2}), where $so(1,2)$ appears as well, the new, highly interesting, feature of \eqref{eq16}/\eqref{eq18} is the $\tau$-dependence of the $so(1,2)$ generators; though trivial from the Poisson point of view, just corresponding to a `constant' shift of $\mu$ by $e$, it actually implies the existence of the following underlying structure, following from
%%%%%%%%%%%%%%%%%%%%%%%%%%%%%%%%%%%%%%%%%%%%%%%%%%%%%%%%%%%%%%%%%%%%%%%%%%%%%
%PAGE 7 PAGE 7 PAGE 7 PAGE 7 PAGE 7 PAGE 7 PAGE 7 PAGE 7 PAGE 7 PAGE 7 PAGE 7 
%%%%%%%%%%%%%%%%%%%%%%%%%%%%%%%%%%%%%%%%%%%%%%%%%%%%%%%%%%%%%%%%%%%%%%%%%%%%% 
$h_1 = \cos \psi$, $h_2 = \sin \psi$, $y = \mu$ satisfying the $e(2)$ relations
\begin{equation}\label{eq19} 
\{ h_1, y \} = h_2, \; \{ y, h_2 \} = h_1, \; \{ h_1, h_2 \} = 0,
\end{equation}
and $h_1^2 + h_2^2 = 1$.
\begin{equation}\label{eq20} 
\bar{x}_1 = (y + e)h_1, \; \bar{x}_2 = (y + e)h_2, \; \bar{x}_0 = y + e
\end{equation}
with
\begin{equation}\label{eq21} 
\bar{x}^2_0 - \bar{x}^2_1 - \bar{x}^2_2 = 0
\end{equation}
then necessarily satisfy \eqref{eq18}, for any (central) $e$. \\
This clearly suggests to start with 3 elements $H_1$, $H_2$, and $Y$ satisfying
\begin{equation}\label{eq22} 
\begin{array}{l}
[H_1, Y] = iH_2, \; [Y, H_2] = iH_1,\; [H_1, H_2] = 0\\[0.15cm]
H^2_1 + H^2_2 = \mathbb{1} 
\end{array}
\end{equation}
(i.e. a `fuzzy cylinder' \cite{6}) and define, as non-commutative analogues of \eqref{eq20}
\begin{equation}\label{eq23} 
\begin{split}
X_1 & := \frac{1}{2}(H_1Y + YH_1) + eH_1 =: Y_1+eH_1\\
X_2 & := \frac{1}{2}(H_2Y + YH_2) + eH_2 =: Y_2+eH_2\\
X_0 & := Y + e\mathbb{1} =: Y_0 + eH_3.
\end{split}
\end{equation}
It is straight forward to deduce (alone from \eqref{eq22}) that both the $X_{\alpha}$ {\it and} the $Y_{\alpha}$ are representations of $so(1,2)$, i.e.
%%%%%%%%%%%%%%%%%%%%%%%%%%%%%%%%%%%%%%%%%%%%%%%%%%%%%%%%%%%%%%%%%%%%%%%%%%%%%
%PAGE 8 PAGE 8 PAGE 8 PAGE 8 PAGE 8 PAGE 8 PAGE 8 PAGE 8 PAGE 8 PAGE 8 PAGE 8  
%%%%%%%%%%%%%%%%%%%%%%%%%%%%%%%%%%%%%%%%%%%%%%%%%%%%%%%%%%%%%%%%%%%%%%%%%%%%% 
\begin{equation}\label{eq24} 
[X_{\alpha}, X_{\beta}] = i\varepsilon_{\alpha \beta} \,^{\gamma}X_{\gamma},\; 
[Y_{\alpha}, Y_{\beta}] = i\varepsilon_{\alpha \beta} \,^{\gamma}Y_{\gamma},
\end{equation}
noting/verifying in particular
\begin{equation}\label{eq25} 
\begin{split}
[Y_{\alpha}, H_{\beta}] + [H_{\alpha}, Y_{\beta}] & = i\varepsilon_{\alpha \beta} \,^{\gamma}H_{\gamma} \\[0.15cm]
\eta^{\alpha \beta} (H_{\alpha} Y_{\beta} + Y_{\alpha}H_{\beta}) & = 0 \\[0.15cm]
H_1 Y H_1 + H_2 Y H_2 & = Y \\[0.15cm]
H_1 Y ^2 H_1 + H_2 Y^2 H_2 & = Y^2 + \mathbb{1};
\end{split}
\end{equation}
while $\hat{X}_1 = \frac{\sqrt{2}}{\tau}X_1(\tau)$ and $\hat{X}_2 = \frac{\sqrt{2}}{\tau}X_2(\tau)$ then automatically satisfy \eqref{eq1} (the `quantum' - calculation trivially following the classical one) there is one significant quantum reflection of the classical singularity\footnote{in \cite{7} non-commutative analogues $\hat{\sum}(t)$ of a 1-parameter family of compact 2 dimensional surfaces $\sum(t)$ going through a singularity, $\sum(t_0)$, were found ($\sum_{t>t_0}$ being spheres, topologically, and $\sum_{t<t_0}$ being tori) with $\hat{\sum}(t_0)$ being `perfectly fine', {\it except} exhibiting a sudden change in representation/dimension of the finite dimensional matrices representing $\hat{\sum}(t)$. As in \cite{2}, for a family of time-like regular faces in $\mathbb{R}^{1,3}$, a perfect match was found between classical and quantum Casimirs, it is intriguing to attribute the change form $0$ to $\frac{1}{4}$ (cp.\eqref{eq26}) to a reflection of singularities of \eqref{eq4} (presumably the sharp edge, possibly the light-like line)}, 
namely the change of Casimir-value(s) from zero (cp.\eqref{eq21}) to
\begin{equation}\label{eq26} 
\begin{split}
-Q & := Y_0^2 - Y_1^2 - Y_2^2 = -\frac{1}{4}\mathbb{1}\\
-Q'& := X_0^2 - X_1^2 - X_2^2 \\
& = (Y_0 + e)^2 - (Y_1 + eH_1)^2 - (Y_2 + eH_2)^2 = -\frac{1}{4}\mathbb{1}.
\end{split} 
\end{equation}
%%%%%%%%%%%%%%%%%%%%%%%%%%%%%%%%%%%%%%%%%%%%%%%%%%%%%%%%%%%%%%%%%%%%%%%%%%%%%
%PAGE 9 PAGE 9 PAGE 9 PAGE 9 PAGE 9 PAGE 9 PAGE 9 PAGE 9 PAGE 9 PAGE 9 PAGE 9 
%%%%%%%%%%%%%%%%%%%%%%%%%%%%%%%%%%%%%%%%%%%%%%%%%%%%%%%%%%%%%%%%%%%%%%%%%%%%% 
What about (related to Lorentz-invariance, and classically necessary to reconstruct the wordvolume)\footnote{the last term, put in with hindsight, to cancel the otherwise occurring discrepancy between $\ddot{\hat{\zeta}}$ and $\hat{\Delta}\hat{\zeta}$ (which inherits an `anomalous term from the right-hand side of \eqref{eq26}) is a pure quantum effect; because of the numerical value (the factor $\frac{1}{12}$), and having in mind the general `reconstruction-algebra' (generalizing the Virasoro algebra to arbitrarily extended objects) that was discovered in \cite{8} it is tempting to speculate that there is a central-extension interpretation to it; in any case the existence of $\hat{\zeta}$ satisfying \eqref{eq27} is highly significant, as it is linked to the issue of Lorentz invariance of the GH-BFSS matrix model (which most people believe to/be proven to/not be realizable for \eqref{eq1}; I disagree/am mildly optimistic)}
\begin{equation}\label{eq27} 
\begin{split}
\hat{\zeta} &\stackrel{?}{=} \varepsilon^2 \tau^3 \mathbb{1} - \frac{Y^2}{\tau^3} + 2\sqrt{2} \varepsilon Y - \frac{\mathbb{1}}{12\tau^3} \\
\hat{\Delta} \hat{\zeta} & = \frac{2}{\tau^2} \Delta (2\sqrt{2}\varepsilon Y - \frac{Y^2}{\tau^3})\\
& = -12\frac{Y^2}{\tau^5} + 6\varepsilon^2 \tau -\frac{1}{\tau^5} \\[0.15cm]
\ddot{\hat{\zeta}} & = 6\varepsilon^2 \tau - 12\frac{Y^2}{\tau^5} - \frac{1}{\tau^5}\\[0.15cm]
\hat{\square} \hat{\zeta} & = (\partial^2_t - \hat{\Delta)}\hat{\zeta} = 0;\\
\end{split}
\end{equation}
which, using \eqref{eq24} and \eqref{eq22}, indeed follows from
\newpage
\begin{equation}\label{eq28}
\begin{split}
\Delta Y & := -\big[ [Y, X_j], X_j \big] = -\big[ [X_i, X_j], X_j \big]\\[0.15cm] 
& = -i \varepsilon_{0j}\,^k [X_k, X_j] = \varepsilon_{0j}\,^k \varepsilon_{kj}\,^0 X_0\\[0.15cm]
& = 2(Y + e\mathbb{1}) \\[0.15cm]
\Delta Y^2 & := -\big[ Y[X_0, X_j], X_j \big] - \big[ [X_0, X_j]Y, X_j \big] \\[0.15cm]
& = Y \Delta Y + (\Delta Y)Y -2[X_0, X_1]^2 - 2[X_0, X_2]^2 \\[0.15cm]
& = 4Y(Y + e) + \underbrace{2X_2^2  +2 X_1^2}_{2(X_0 = Y+e)^2 +2/4} \\
& = 6Y^2 + 8Ye + 2e^2 + \frac{1}{2}.\\[0.15cm]
\end{split}
\end{equation}
%%%%%%%%%%%%%%%%%%%%%%%%%%%%%%%%%%%%%%%%%%%%%%%%%%%%%%%%%%%%%%%%%%%%%%%%%%%%%
%PAGE 10 PAGE 10 PAGE 10 PAGE 10 PAGE 10 PAGE 10 PAGE 10 PAGE 10 PAGE 10 PAGE 10  
%%%%%%%%%%%%%%%%%%%%%%%%%%%%%%%%%%%%%%%%%%%%%%%%%%%%%%%%%%%%%%%%%%%%%%%%%%%%% 
It is instructive to also check  by a {\it direct} calculation that the quantized embedding coordinates $\hat{X}_1$, $\hat{X}_2$ and $\hat{\zeta}$ are annihilated by the ($\hat{X}_i$-dependent) quantum wave operator $\square$, by taking concrete representations of \eqref{eq22}, resp. (letting $W := H_1 + iH_2$, $W^\dagger = H_1 - iH_2$)
\begin{equation}\label{eq29} 
Y|n> = (n+\kappa)|n>, \quad W|n> = , \quad W^\dagger|n> = |n+1>
\end{equation}
$n \in \mathbb{Z}$, $\kappa \in [0,1)$, note that $WW^\dagger = 1 = W^\dagger W$
\begin{equation}\label{eq30} 
\begin{split}
Z & := X_1 + i X_2 = \frac{1}{2}(WY + YW) + eW\\
Z|n> & = Z_n|n-1>, \quad Z^\dagger |n>  = Z_{n+1}|n+1>\\
Z_n  & = n - \frac{1}{2} + e + \kappa.
\end{split}
\end{equation}
\begin{equation}\label{eq31} 
-  \big[ [X_j, X_k], X_k \big] = X_j \Leftrightarrow  \big[ [Z, Z^\dagger], Z \big] = -2Z, 
\end{equation}
acting an $|n>$ gives $2Z^3_n - Z^2_{n-1}Z_n - Z_nZ^2_{n+1} = -2Z_n$ i.e. (for $Z_n \neq 0$) the condition
\begin{equation}\label{eq32} 
2Z^2_n - Z^2_{n-1} - Z^2_{n+1} = -2, 
\end{equation}
which {\it is} satisfied for \eqref{eq30} (any $e$ and $\kappa$); in fact $Z_n = \pm n +${\it any} constant will work,
%%%%%%%%%%%%%%%%%%%%%%%%%%%%%%%%%%%%%%%%%%%%%%%%%%%%%%%%%%%%%%%%%%%%%%%%%%%%%
%PAGE 11 PAGE 11 PAGE 11 PAGE 11 PAGE 11 PAGE 11 PAGE 11 PAGE 11 PAGE 11 PAGE 11 
%%%%%%%%%%%%%%%%%%%%%%%%%%%%%%%%%%%%%%%%%%%%%%%%%%%%%%%%%%%%%%%%%%%%%%%%%%%%% 
resp. (general solution of the simple recursion relation \eqref{eq32})
\begin{equation}\label{eq33} 
Z_n = \pm \sqrt{n^2 + 2\gamma n + \delta}.
\end{equation}
Verifying \eqref{eq28} on the other hand is more involved : %assuming $\hat{\zeta}|n> = \zeta_n|n>$ 
\begin{equation}\label{eq34} 
\begin{split}
\Delta \hat{\zeta} & = - \big[ [\hat{\zeta}, X_i], X_i \big]\\
& = -\frac{1}{2} \big[ [\hat{\zeta}, Z], Z^\dagger \big]  -\frac{1}{2} \big[ [\hat{\zeta}, Z^\dagger], Z \big]\\
& = Z \zeta Z^\dagger + Z^\dagger \zeta Z - \frac{1}{2} \zeta(ZZ^\dagger + Z^\dagger Z) -\frac{1}{2} (ZZ^\dagger + Z^\dagger Z)\zeta
\end{split}
\end{equation}
gives ( assuming $\hat{\zeta}|n> = \zeta_n|n>$ )
\begin{equation}\label{eq35} 
(\Delta \hat{\zeta})_n = Z^2_{n+1}(\zeta_{n+1}-\zeta_n) - Z^2_n (\zeta_n - \zeta_{n-1});
\end{equation}
\begin{equation}\label{eq36} 
\begin{array}{l}
Z^2_n(\zeta_n-\zeta_{n-1}) = \alpha n^3 + \beta n^2 + \gamma n + \delta \\[0.15cm]
\alpha = \frac{-2}{\tau^3}, \; \beta = \frac{3}{\tau^3}, \gamma = 3 \varepsilon^2 \tau^3 - \frac{3}{2\tau^3}, 
\end{array}
\end{equation}
here taking $\kappa = 0$ for simplicity, gives
\begin{equation}\label{eq37} 
\begin{split}
(\Delta \hat{\zeta})_n & = 3\alpha n^2 + n(3\alpha + 2\beta) + (\alpha + \beta + \gamma)\\
& = -\frac{6n^2}{\tau^3} + 0 +3\varepsilon^2 \tau^3 - \frac{1}{2\tau^3}, 
\end{split}
\end{equation}
which matches $\frac{\tau^2}{2} \ddot{\zeta}_n = \frac{\tau^2}{2}(6\varepsilon^2 \tau - \frac{12 n^2}{\tau^5} + 0 -\frac{1}{\tau^5})$; note that without the quantum effect (the last term in \eqref{eq27}) \eqref{eq37} (the last term arising from \eqref{eq36}, no matter what) would {\it not} be equal to $\frac{\tau^2}{2}\ddot{\zeta}_n$.\\
\textbf{Acknowledgement}:
I would like to thank J.Arnlind, T.Damour, J.Eggers, M.Kontsevich, V.Roubtsov and V.Sokolov for discussions.


\begin{thebibliography}{11111}
\bibitem[1]{1}  J.Hoppe, {\it Quantum Theory of a Massless Relativistic Surface}, MIT Ph.D. Thesis 1982,
http://dspace.mit.edu/handle/1721.1/15717
\bibitem[2]{2} J.Hoppe, {\it Recent Progress on Membrane Theory}, Proceedings of Science, PoS(Corfu2021)258
\bibitem[3]{3} J.Hoppe, {\it Some classical solutions of relativistic membrane equations in 4 space-time dimensions}, Phys.Lett.B329, 1994 66
\bibitem[4]{4} J.Hoppe, {\it On the quantization of some polynomial minimal surfaces}, Phys.Lett.B822 2021
\bibitem[5]{5} J.Hoppe, Gauge-compensating transformations for boosted axially symmetric membranes and light-cone reductions 
( manuscript,  2023)
\bibitem[6]{6} M.Chaichian, A.Demichev, P.Presnadjer, {\it Quantum field theory on non-commutative space-times and the persistence of ultraviolet divergences}, Nucl.Phys.B567(2000)
\bibitem[7]{7} J.Arnlind, M.Bordemann, L.Hofer, J.Hoppe, H.Shimada, {\it Fuzzy Riemann Surfaces}, JHEP06(2009)047
\bibitem[8]{8} J.Hoppe, {\it Fundamental Structures of M(brane) theory}, Phys.Lett.B695 2011
\end{thebibliography}
\end{document}